\documentclass[a4paper,10pt]{article}
\usepackage{amsmath}
\usepackage{graphicx}
\usepackage{amssymb}
\title{Classical Correlations and Quantum Interference in Ballistic Conductors}
\author{Daniel Waltner and Klaus Richter\vspace*{3mm}\\Institut f\"ur Theoretische Physik\\  Universit\"at Regensburg, D-93040 Regensburg, Germany}

\begin{document}

\maketitle
\begin{abstract}
We illustrate how classical chaotic dynamics influences the quantum 
properties at mesoscopic scales. As a model case we study semiclassically coherent 
transport through ballistic mesoscopic systems 
within the Landauer formalism beyond the so-called diagonal approximation, \textit{i.e.}\ by incorporating  classical action correlations. In this 
context we review and explain the two main trajectory-based methods developed for calculating quantum corrections: the configuration 
space approach and the phase space approach that can be regarded as less illustrative but more general than 
the first one.

\end{abstract}

\section{Introduction: quantum transport \\
         through chaotic conductors}

Among the most important properties characterizing an electronic nanosystem is 
its electrical conductance behavior. Hence gaining knowledge on charge transport
mechanisms, in particular when shrinking conductors from macroscopic sizes
down to molecular-sized wires or atomic point contacts, has been in the
focus of experimental and theoretical research throughout the last decade.
Such a reduction in size and spatial dimensionality goes along with a crossover
from charge flow in the macroscopic bulk, well described by Ohm's law, to distinct
quantum effects in the limit of microscopic or atomistic wires. Nanoconductors
in the crossover regime, often referred to as mesoscopic, frequently exhibit a
coexistence of both classical remnants of bulk features, combined with signatures from wave
interference. Such quantum effects usually require low temperatures where coherence
of the electronic wavefunctions is retained up to micron scales. This has lead to
the observation of various quantum interference phenomena such as quantized
steps in the point contact conductance, the Aharonov-Bohm effect and universal
conductance fluctuations, to name only a few. 

Nonlinear effects can enter into transport through mesoscopic or nano-systems twofold:
First, as nonlinear I-V characteristics and charge flow far from equilibrium for large 
enough voltage applied. Second, in the limit of linear response to an applied 
electric field, the intrinsic nonlinear classical dynamics of the unperturbed conductor can 
govern its transport properties. In this chapter we focus on the latter case which
is particularly interesting for mesoscopic conductors because the nonlinear charge
carrier dynamics can influence both the classical and, in a more subtle way, the
quantum transport phenomena.

Initially, disordered metals with underlying diffusive charge carrier motion were 
in the focus of interest in mesoscopic matter. Here, we instead address ballistic nano-
or mesoscopic conductors where impurity scattering is suppressed. The most prominent
ballistic systems are nanostructures built from high-mobility semiconductor heterostructures,
where electrons are confined to two-dimensional, billiard-type cavities of controllable
geometry; however, such systems are also realized as atom optics billiards or through
wave scattering as optical, microwave, or acoustic  mesoscopic resonators
\cite{Stoeckmann}.

\begin{figure}
\centerline{\includegraphics[width=0.9\textwidth]{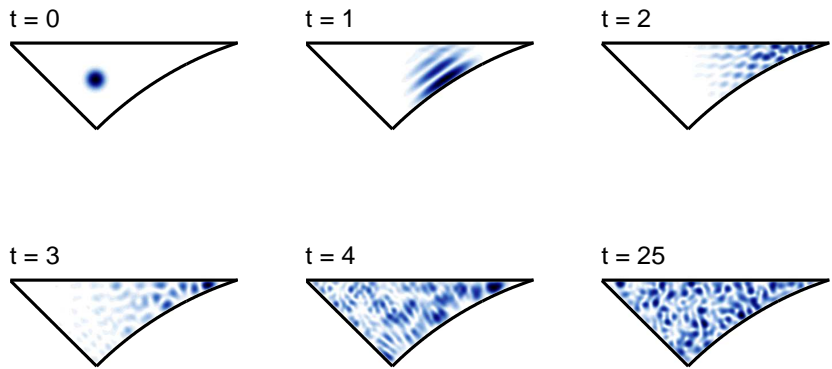}}

\caption{Quantum mechanical wave packet launched into a mesoscopic cavity with
the geometry of a "desymmetrized diamond billiard". The wave packet evolution
is monitored at times $t=$  1, 2, 3, 4, and 25, in units of the average time between
collisions  with the walls of a corresponding classical particle (Courtesy of A.\ Goussev).}
\label{fig:wavepacket}
\end{figure}

The quantum dynamics of a single particle in such a geometry is illustrated 
in Fig.\ \ref{fig:wavepacket} showing snapshots of a wave packet after multiples
of the average classical time between bounces off the billiard walls. The quantum
evolution in such a mesoscopic geometry, with corresponding chaotic classical
dynamics, is characterized by two main features: The rapid transition from wave
packet motion following roughly the path of a classical particle to random wave
interference at larger times and, second, the emergence of wave functions of complex morphology
with wave lengths much shorter than the system size. The latter can be 
used to further specify mesoscopic matter, {\em i.e.}\ quantum coherent systems where
the smallest (quantum) length scale, the de Broglie wave length or Fermi wave length 
$\lambda_{\rm F}$ in electronic conductors, is much smaller than the system size 
$\cal L$, {\em i.e.}\ $1/(k_{\rm F} \cal L)$ is a small parameter, in terms of the Fermi
momentum $k_{\rm F} = 2\pi / \lambda_{\rm F}$, but not fully negligible as
in the case of macroscopic systems.  

Such ballistic mesoscopic systems are ideal tools to study the connection between 
(chaotic) classical dynamics and wave interference. Semiclassical techniques provide 
this link presumably in the most direct way.  Modern semiclassical theory is based on 
trace formulas, sums over Fourier-type components associated with classical trajectories.
Analogous to the famous Gutzwiller trace formula for the density of states \cite{gutz_book}, 
there exist corresponding expressions for quantum transport in the linear response regime.
There, semiclassical expressions for the conductance have been obtained within the
framework of the Landauer-B\"uttiker approach, relating conductance to quantum
transmission in nanostructures.

After early, pioneering semiclassical work by Miller \cite{Miller74} for molecular 
reactions and later by Bl\"umel and Smilansky \cite{Bluemel88} for quantum 
chaotic scattering, major advances were made in the context of mesoscopic conductance 
in the early nineties by Baranger, Jalabert and Stone \cite{BJS90,Baranger93}.
All these semiclassical approaches were based on, and limited by, the so-called diagonal 
approximation, see below. While most of the features of experimental and numerical magneto-conductance profiles could be well explained qualitatively on the level of the diagonal 
approximation, it contained the major drawback that it was not current-conserving and hence 
failed to give correct quantitative predictions for the quantum transmission. 
This was cured about 10 years later when an approach was devised to account
for off-diagonal contributions to the semiclassical conductance \cite{Richter02},
thereby achieving unitarity, reflected in (average) current conservation, and furthermore 
agreement with existing predictions from random matrix theory (RMT). The applicability of RMT to closed chaotic mesoscopic systems was conjectured after numerical simulations in \cite{Boh}. 
So it was expected to be applicable also to open systems.

Semiclassical ballistic transport on the level of the diagonal approximation was 
reviewed in detail in Refs.\ \cite{BJS93,Richter00,Jalabert00}. In this chapter we hence
focus on recent progress beyond the diagonal approximation during the last years. This
serves as a model case which illustrates how chaotic nonlinear dynamics can govern quantum 
properties at nanoscales.

\section{Semiclassical limit of the Landauer transport approach}

The Landauer formalism \cite{Datta}, providing a link between the quantum transmission and conductance,
has proven to be an appropriate framework to address phase-coherent transport through nanosystems.
Consider a sample attached to two leads of width $W_1$ and $W_2$ that support $N_1$, respectively,
$N_2$ current carrying transverse modes at (Fermi-)energy $E_F$. For such a two-terminal setup the
conductance reads at very low temperatures \cite{Datta}
\begin{equation}
\label{Landauer}
G(E_F) = g_{\rm s} \frac{e^2}{h} T(E_F)  = g_{\rm s}  \frac{e^2}{h} \sum_{m=1}^{N_1} \sum_{n=1}^{N_2} |t_{nm}(E_F)|^2 \; 
\end{equation}
with $N_1=W_1\sqrt{2mE_F}/(\hbar\pi)$ and an analogous relation for $N_2$. Here, $g_{\rm s} = 2$  accounts for spin degeneracy, and the $t_{nm}(E)$ are transmission amplitudes 
between incoming channels $m$
and outgoing channels $n$ in the leads at energy $E$.  They can be expressed in terms of the projections 
of the Green function of the scattering region onto the transverse modes 
$\phi_n(y')$ and $\phi_m(y)$ in the two leads
\cite{Fisher-Lee}:
\begin{eqnarray}
\label{Greenproj}
t_{nm}(E)  = - {\rm i} \hbar(v_n v_m)^{1/2}
\int {\rm d}y' \int
{\rm d}y\,\phi_n^\ast(y') \phi_m(y) \ G(x',y',x,y;E)\; .
\end{eqnarray}
Here $x$ and $x'$, respectively, denote the direction along the leads and 
$v_n, v_m$ the corresponding longitudinal velocities. The integrals in Eq.\ (\ref{Greenproj})
are taken over the cross sections of the (straight) leads at the entrance and the exit.

\begin{figure}[ht]
\centerline{
\includegraphics[width=.85\textwidth]{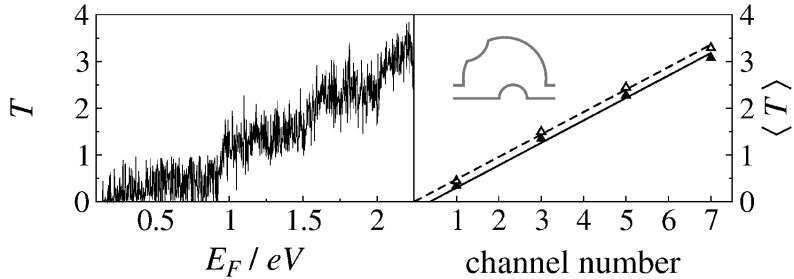}
}
\caption{Total quantum transmission (as a function of energy $E_F$,
on the right in units of the channel number in the leads of width $W_1=W_2=W$)
for transport through a phase-coherent graphene-based quantum dot, see inset.
The fluctuating line in the left panel is the full quantum transmission
at zero magnetic field. In the right panel, the straight solid  and dashed line 
denote the averaged transmission at zero magnetic field and a magnetic field corresponding 
to a flux $\phi = 1.6 \phi_0$ with the flux quantum $\phi_0=hc/e$. The difference marks 
the weak localization correction (from Ref.~\cite{Wurm08}).}
\label{fig:wurm}
\end{figure}

Fig.\ \ref{fig:wurm} shows the  quantum transmission, numerically obtained from Eq.\ (\ref{Greenproj}),
for a "graphene billiard" \cite{Wurm08}, fabricated by cutting a cavity out of a two-dimensional graphene
flake, a monoatomic layer of carbon atoms arranged in a honeycomb lattice.
 Two major quantum features are visible:
(i) distinct "ballistic conductance fluctuations" as a function of energy. 
(ii) When subject to an additional, perpendicular magnetic field $B$ with
magnetic flux $\phi$, the average transmission (straight dashed line in the right panel) 
shows a small positive offset
compared to the average transmission for $B=0$  (solid line). This reduction of the average
conductance at zero magnetic field reflects a weak-localization effect \cite{CS86}. Its origin is 
non-classical and due to wave interference. 

Here we focus on this ballistic weak localization effect
and present its semiclassical derivation for conductors with classically chaotic analogue.
The semiclassical approximation enters in two steps: First, we replace $G(x',y',x,y;E)$ 
in Eq.\ (\ref{Greenproj}) by the semiclassical Green function (in two dimensions)
\cite{gutz_book}:
\begin{equation}
 \label{eq:green}
  G^{\rm sc}({\bf r}',{\bf r};E)
               =
            \frac{1}{{\rm i}  \hbar(2 {\rm i}  \pi \hbar)^{1/2} }
   \sum_t D_t({\bf r}', {\bf r}) \exp{\left(\frac{{\rm i} }{\hbar} S_t - 
   {\rm i} \eta_t\frac{\pi}{2}\right)} . 
 \end{equation}
It is given as a sum over  contributions 
from all classical trajectories $t$ connecting
the two fixed points ${\bf r}$ and ${\bf r}'$ at energy $E$. In Eq.\ (\ref{eq:green}),
     \begin{equation}
     \label{action}
        S_t({\bf r}',{\bf r};E) = \int_{{\cal C}_t} \vec{p} \cdot {\rm d}  \vec{q}
   \end{equation}
is the classical action along a path ${\cal C}_t$ between ${\bf r}$ and ${\bf r}'$
	and governs the accumulated phase.

 Second, we evaluate  the projection integrals in Eq.\ (\ref{Greenproj}) 
 for isolated trajectories within the stationary-phase approximation.
 For leads with hard-wall boundaries the mode wavefunctions are sinusoidal,
 $\phi_m(y) = \sqrt{2/W_1}\sin(m\pi y/W_1)$. Hence the stationary-phase condition
 for the $y$ integral requires \cite{BJS93}
 \begin{equation}
 \label{statphase}
 \left(\frac{\partial S}{\partial y}\right)_{y'} = -p_y \equiv
 -\frac{\overline{m}\hbar\pi}{W_1} \; ,
 \end{equation}
 with $\overline{m} = \pm m$. The stationary-phase solution of the $y'$ integral
 yields a corresponding ``quantization'' condition for the transverse momentum
 $p_{y'}$.
 Thus only those paths which
 enter into the cavity at $(x,y)$ with a fixed angle $\sin\theta=\pm m\pi/kW_1$ and
 exit the cavity at $(x',y')$ with angle $\sin\theta'=\pm n\pi/kW_2$ contribute 
 to $t_{nm}(E)$ with $k=\sqrt{2mE}/\hbar$. There is an intuitive explanation: the trajectories
 are those whose transverse wave vectors on entrance and exit
 match the wave vectors of the modes in the leads.
One then obtains for the
semiclassical transmission amplitudes\index{transmission amplitude}
\begin{equation}
\label{transamp}
\hspace*{-6mm}
t_{nm}(k)  =   -\frac{\sqrt{2\pi {\rm i}\hbar}}{2\sqrt{W_1W_2}}\! \sum_{t(\overline{n},\overline{m})}
{\rm sgn}(\overline{n})\ {\rm sgn}(\overline{m}) \sqrt{A_t}
       \exp{\left[\frac{{\rm i}} {\hbar} \tilde{S}_t(\overline{n},\overline{m};k) -{\rm i}  \frac{\pi}{2}
                  \tilde{\mu}_t \right]} .
\end{equation}
	  Here, the reduced actions\index{action} are
 \begin{equation}
 \label{modaction}
		  \tilde{S}_t(\overline{n},\overline{m};k) = S_t(k)
		  + \hbar k y \sin\theta-\hbar ky'\sin\theta',
\end{equation}
   which can be considered as Legendre transforms of the original action functional.
   The phases $\tilde{\mu}_t$ contain both the usual Morse indices and additional
   phases arising from the $y,y'$ integrations. The prefactors are
   $A_t = |\left( \partial y/\partial \theta'\right)_\theta |/\left(\hbar k \left|\cos \theta'\right|\right)$.
    The resulting semiclassical expression for the
    transmission and thereby
  the conductance\index{conductance, semiclassical approximation}
 (see Eq.\ (\ref{Landauer})) in chaotic cavities involves contributions
  from pairs of trajectories $t,t'$. It reads \cite{BJS90,Baranger93,BJS93}

\begin{equation}
\label{Gsk}
T(k) = \sum_{m=1}^{N_1} \sum_{n=1}^{N_2} |t_{nm}(k)|^2 =
  \frac{\pi\hbar}{2{W_1W_2}} \sum_{m=1}^{N_1} \sum_{n=1}^{N_2} \sum_{t,t'} \ F_{n,m}^{t,t'}(k) \; 
\end{equation} 
  with
\begin{equation}
  \label{Fnm}
  F_{n,m}^{t,t'}(k) \equiv  \sqrt{A_t A_{t'}} \ \exp{\left[\frac{{\rm i} }{\hbar}(\tilde{S}_t-
                     \tilde{S}_{t'}) - {\rm i} \mu_{t,t'}\frac{\pi}{2}\right]} \; ,
\end{equation} 
where the phase $\mu_{t,t'}$ accounts for the differences of the phases $\tilde\mu_t$ and the sgn
factors in Eq.\ (\ref{transamp}).

In the next two sections we show how one can calculate with purely semiclassical methods
quantum corrections to the transmission beyond the diagonal approximation.
The results obtained for chaotic conductors 
are consistent with RMT-predictions.

\section{Quantum transmission: \\ 
           configuration space approach}
\label{sec:config-space}

In evaluating the off-diagonal (interference) contributions to the quantum transmission
we present two approaches. The first approach is based on the analysis of off-diagonal pairs
of trajectories and their self-intersections in configuration space. This approach is
more illustrative, but less general
than the second, phase space approach outlined in Sec.~\ref{sec:phase-space}.

We start from the semiclassical expression for $\left|t_{nm}(k)\right|^2$ obtained by
squaring the semiclassical transmission amplitudes, Eq.\ (\ref{transamp}), see also Eqs.\ (\ref{Gsk},\ref{Fnm}):
\begin{equation}
\left|t_{nm}(k)\right|^{2}=\frac{\pi\hbar}{2W_1W_2}\sum_{t,t'}\sqrt{A_t A_{t'}} \ \exp{\left[\frac{{\rm i} }{\hbar}(\tilde{S}_t-
                     \tilde{S}_{t'}) -  {\rm i} \mu_{t,t'}\frac{\pi}{2}\right]}.\label{eq:tra1}
\end{equation}
%
Due to the action difference of the trajectories $t$ and $t'$ in the exponential in Eq.\ (\ref{eq:tra1}), 
this expression is a rapidly oscillating function of $k$ or the energy in the semiclassical limit of
large ratios $(S_t-S_{t'}) / \hbar$. In the following we wish to identify those contributions to 
Eq.\ (\ref{eq:tra1}) which survive  an average over a classically small but quantum mechanically 
large $k$-window $\Delta k$. Such contributions must come from very similar trajectories. Afterwards we
wish to evaluate their contributions to $\left|t_{nm}(k)\right|^2$ using basic principles of chaotic
dynamics: for our calculation we will need hyperbolicity and ergodicity. Hyperbolicity implies the
possible exponential separation of neighboring trajectories for long times with the distance growing
proportional to $e^{\lambda T}$,
with the Lyapunov exponent $\lambda$ and the time $T$ of the trajectory. 
The second principle, ergodicity, means the equidistribution of long trajectories on the energy surface
at the energy $E$ of the trajectory.

This section is divided into four subsections: after calculating the diagonal contribution  
we determine the simplest nondiagonal contribution in the second part. The behavior of the latter 
as a function of a magnetic field and a finite Ehrenfest time is then studied in the third and fourth 
subsection, respectively.

\subsection{Diagonal contribution}

The first, "diagonal" (D)
contribution to Eq.\ (\ref{eq:tra1}) originates from identical trajectories $t=t'$, meaning $S_t=S_{t'}$. It gives 
\begin{equation}
\left|t_{nm}(k)\right|^{2}_D=\frac{\pi\hbar}{2W_1W_2}\sum_t A_t.
\label{dia1} 
\end{equation}
The remaining sum over classical trajectories in Eq.\ (\ref{dia1}) can be calculated using a classical 
sum rule \cite{Richter02} that can be derived using ergodicity, see for example \cite{Sieb}. It yields
\begin{equation}
\sum_t A_t=\frac{4W_1W_2}{\Sigma(E)}\int_0^\infty\rho(T) , 
\label{dia2}
\end{equation}
where $\Sigma(E)$ denotes  the phase space volume of the system at energy $E$, and  $\rho(T)$ is
the classical probability to find a particle still inside an open system after a time $T$. 
For long times the latter decays for a chaotic system exponentially $\rho(T)\sim e^{-T/\tau_D}$,
with the dwell time $\tau_D=\frac{\Sigma(E)}{2\pi\hbar\left(N_1+N_2\right)}$. 
This exponential decay can be easily understood based on the equidistribution of trajectories: 
the number $\Delta N$ of particles leaving the system during $\Delta T$ is given by the overall 
number of particles $N$ times the ratio of the phase space volume from which the particles 
leave during $\Delta T$ and the whole phase space volume of the system. The differential equation for
$N$, obtained in the case of infinitesimal $\Delta T$, has obviously an exponential solution. 

By inserting  Eq.\ (\ref{dia2}) into Eq.\ (\ref{dia1}), we obtain
\begin{equation}
\left|t_{nm}(k)\right|^{2}_D=\frac{1}{N_1+N_2} \, .
\label{dia3}
\end{equation}
Its derivation required ergodicity valid only for long trajectories. We will assume that the 
classical dwell time is large enough, in order to have a statistically relevant number of 
long trajectories left after time $t$.  The result in Eq.\ (\ref{dia3}) allows for a very simple 
interpretation: It is just the probability of reaching one of the $N_1+N_2$ channels if each of 
the channels can be reached equally likely.

\subsection{Nondiagonal contribution}

In the following we will calculate the contributions from pairs of different trajectories, however
with similar actions. For a long time it was not clear how these orbits could look like. There are many orbit pairs in a chaotic 
system having accidentally equal or nearly equal actions. However, in order to describe universal features 
of a chaotic system after energy averaging, one has to find orbits that are correlated in a systematic way.
These orbits were first identified and analyzed in 2000 in the context of spectral
statistics \cite{Sie}. There, {\em periodic} orbits were studied to compute correlations between energy 
eigenvalues of quantum systems with classically chaotic counterpart. Based on {\em open},
lead-connecting trajectories, in Ref.~\cite{Richter02} this approach was generalized to the conductance 
we study here.  Still, the underlying mechanism to form pairs of classically correlated orbits is
the same in the two cases. 

\begin{figure}
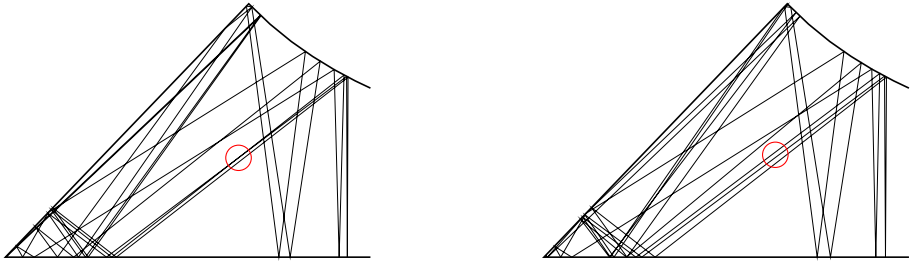


\includegraphics[width=0.4\textwidth]{waltner-richter-fig3a.eps}
\hspace*{2cm} 
\includegraphics[width=0.4\textwidth]{waltner-richter-fig3b.eps}
\caption{
Pair of two periodic orbits in the hyperbola billiard differing from each other essentially
in the region marked by the circle, where the left orbit exhibits a self-crossing while the
right partner orbit shows an "avoided crossing"
 (Courtesy of M.\ Sieber).}
\label{fig:orbits}
\end{figure}

\begin{figure}
\centerline{\includegraphics[width=0.6\textwidth]{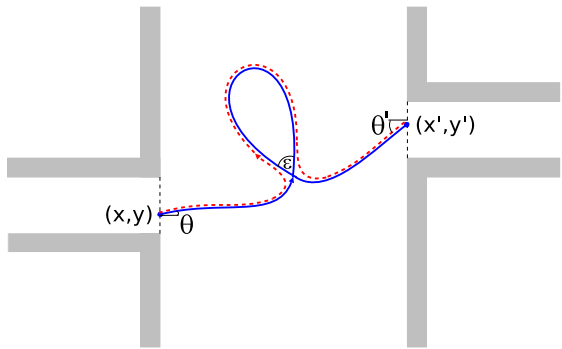}}
\caption{
Schematic drawing of a pair of orbits yielding the first nondiagonal contribution to the transmission considered in Ref.\ \cite{Richter02}. One of the orbits crosses itself under an angle $\epsilon$, the other one possesses an "avoided crossing". Except for the crossing region both orbits are almost identical.
 }
\label{fig:orb}
\end{figure}

In Fig.\ \ref{fig:orbits} we show a representative example of such a correlated (periodic) orbit
pair in the chaotic hyperbola billiard. The two partner orbits are topologically the same up to
the region marked by the circle where one orbit exhibits a self-crossing (left panel) 
while the partner orbit an `avoided' crossing (right panel). Usually, such trajectory pairs
are drawn schematically as shown in Fig.\ \ref{fig:orb}.

One considers very long orbits with self-crossings characterized by crossing angles $\epsilon\ll\pi$. 
In Ref.~\cite{Sie} it was shown 
that there exists for each orbit a partner orbit starting and ending (exponentially) close to the first one. 
It follows the first orbit until the crossing, avoids this, however, traverses the loop in reversed 
direction and avoids the crossing again. 

In order to quantify the contribution of these trajectory pairs to Eq.\ (\ref{eq:tra1}) we need two inputs: an expression for the action difference and for the density quantifying how often an orbit of 
time $T$ exhibits a self-intersection, both quantities expressed as a function of the parameter $\epsilon$. 
The formula for the action difference $\Delta S$ can be derived by 
linearizing the dynamics of the orbit without crossing around the reference orbit with crossing, 
giving, in the limit $\epsilon\ll \pi$, \cite{Sie}
\begin{equation}
\Delta S=\frac{p^{2}\epsilon^{2}}{2m\lambda} \, .
\label{eq:tra10}
\end{equation} 
%

%
%
At this point we can justify our assumption of small crossing angles $\epsilon$:
In the limit $\hbar\rightarrow 0$, we expect important contributions to Eq.\ (\ref{eq:tra1}) 
only from orbit pairs with small action differences, {\em i.e.}\ small crossing angles, 
as we see from Eq.\ (\ref{eq:tra10}).

Before deriving the number of self-crossings, $P\left(\epsilon,T\right)d\epsilon$,
in the range between $\epsilon$ and $\epsilon+d\epsilon$ of an orbit of time $T$,
we give rough arguments how this expression depends on $\epsilon$ and $T$ for trajectories in billiards. There,
each orbit is composed of a chain of $N$ chords connecting the reflection points. Following an orbit,
the first two chords cannot intersect, the third chord can cross with up to one, the fourth chord
with up to two segments, and so on. Hence, the overall number of self-crossings will be proportional to
$\sum_{n=3}^N (n-2) \propto N^2$, to leading order in $N$, {\em i.e.}\ proportional to $T^2$.

The crossing angle dependence of $P\left(\epsilon,T\right)$ can be estimated for small $\epsilon$
as follows: Given a trajectory chord of length $L$, a second chord, tilted by an angle  $\epsilon$
with respect to the first one, will cross it inside the billiard (with area of order $L^2$) 
only if the distance between the reflection points of the two chords at the boundary is smaller 
than $L\sin\epsilon$. The triangle formed in the latter case includes a fraction $\sin \epsilon$ of the entire billiard size.  
From this rough estimation we expect $P\left(\epsilon,T\right) \propto T^2 \sin\epsilon $.

More rigorously, the quantity $P\left(\epsilon,T\right)d\epsilon$,
can be expressed for an arbitrary orbit $\gamma$ as \cite{Sie}
\begin{eqnarray}
P\left(\epsilon,T\right)d\epsilon&=&\left\langle \int_{T_{{\rm min}}(\epsilon)}^{T-T_{{\rm min}}\left(\epsilon\right)}dt_{l}\int_{T_{{\rm min}}(\epsilon)/2}^{T-t_l-{T_{{\rm min}}(\epsilon)/2}}dt_s\left|J\right|\delta\left(\mathbf{x}\left(t_{s}\right)-\mathbf{x}\left(t_{s}+t_{l}\right)\right)\right.\nonumber\\&&\left. \times\delta\left(\epsilon-\alpha\left(t_{s},t_{s}+t_{l}\right)\right)\right\rangle d\epsilon
\label{eq:tra11}
\end{eqnarray}
with the average $\left\langle\ldots \right\rangle$  taken over different initial conditions
$\left(\mathbf x_0,\mathbf p_0\right)$. The time of the closed loop of the trajectory is denoted by
$t_l$ and the time before the loop by $t_s$. $\alpha\left(t_{s},t_{s}+t_{l}\right)$ denotes the absolute
value of the angle between the velocities $\mathbf v \left(t_s\right)$ and $\mathbf v
\left(t_s+t_l\right)$. $\left|J\right|$ is the Jacobian for the transformation from the argument of the
first delta function to $t_l$ and $t_s$ ensuring that $P\left(\epsilon,T\right)d\epsilon$ yields a 1 for
each crossing of $\gamma$. With the absolute value of the velocity, $v$, it can be expressed as 
\begin{eqnarray}
\left|J\right| =\left|\mathbf{v}\left(t_{s}\right)\times\mathbf{v}\left(t_{s}+t_{l}\right)\right|
  =  
v^2\sin\alpha\left(t_{s},t_{s}+t_{l}\right) \, .
\label{eq:tra13}
\end{eqnarray}

As the derivation \cite{Sie} of the formula for $P\left(\epsilon,T\right)$ for a chaotic system,
starting from the formal expression (\ref{eq:tra11}), is instructive to see how information can 
be extracted from the basic principles of chaotic dynamics beyond the diagonal approximation, 
we will present it here in detail. Hyperbolicity will yield a justification for the minimal 
time $T_{{\rm min}}(\epsilon)$ already introduced in Eq.\ (\ref{eq:tra11}); we will come back to 
that point later and first study the effect of ergodicity.

To proceed we interchange the phase space integral of the average with the time integrals, substitute $\left(\mathbf{x}\left(t_{s}\right),\mathbf{p}\left(t_{s}\right)\right)\longmapsto\left(\mathbf{x}_{0},\mathbf{p}_{0}\right)$
in Eq.\ (\ref{eq:tra11}) and obtain 
\begin{eqnarray}
P\left(\epsilon,T\right)  =2m\int_{T_{{\rm min}}\left(\epsilon\right)}^{T-T_{{\rm
min}}\left(\epsilon\right)}dt_{l}v^{2}\sin\epsilon p_{E}\left(\epsilon,t_{l}\right)\left(T-t_{l}-T_{{\rm
min}}(\epsilon)\right)\; ,
\label{eq:tra14}
\end{eqnarray}
with the averaged classical return probability density 
\begin{eqnarray}
p_{E}\left(\epsilon,t_{l}\right)  =\frac{1}{2m}\left\langle  \delta\left(\mathbf{x}_{0}-\mathbf{x}\left(t_{l}\right)\right)\delta\left(\epsilon-\left|\measuredangle\left(\mathbf{v}_{0},\mathbf{v}\left(t_{l}\right)\right)\right|\right)
 \right\rangle \, .
\label{eq:tra15}
\end{eqnarray}
This yields the probability density that a particle possessing the energy $E$ returns after the
time $t_l$ to its starting point with the angle
$\left|\measuredangle\left(\mathbf{v}_{0},\mathbf{v}\left(t_{l}\right)\right)\right|=\epsilon$. For long
times this can be replaced by $1/\Sigma\left(E\right)$, assuming ergodicity. 
Then we obtain 
\begin{eqnarray}
P\left(\epsilon,T\right) & = & 2m\int_{T_{{\rm min}}\left(\epsilon\right)}^{T-T_{{\rm min}}\left(\epsilon\right)}dt_{l}v^{2}\sin\epsilon\frac{1}{\Sigma\left(E\right)}\left(T-t_{l}-T_{{\rm min}}(\epsilon)\right) \nonumber \\&= & \frac{mv^{2}}{\Sigma(E)}\sin\epsilon\left(T-2T_{{\rm min}}\left(\epsilon\right)\right)^2.%
 \end{eqnarray} 

Now we return to our assumption of hyperbolicity and explain the cutoff time $T_{{\rm min}}(\epsilon)$, 
introduced in the equations above. To this end, we consider two classical paths leaving their crossing 
with a small angle $\epsilon$. The initial deviation of their velocities is $\delta v_i=\epsilon v$. 
In order to form a closed loop, the deviation of the velocities
$\delta v_{f}$, when both paths have traversed half of the closed loop, has to be given by $\delta
v_{f}=cv$ with $c$ of the order unity. Then we get for the minimal time $T_{{\rm
min}}\left(\epsilon\right)$ to form a
closed loop, due to the exponential divergence of neighboring orbits,
\begin{equation}
c=\epsilon e^{\left(\lambda T_{{\rm min}}\left(\epsilon\right)\right)/2},
\label{eq:tra20}
\end{equation}
implying
\begin{equation}
T_{{\rm min}}\left(\epsilon\right)=\frac{2}{\lambda}\ln\left(\frac{c}{\epsilon}\right).
\label{eq:tra21}
\end{equation} 
An argument similar to the one used here for the closed loop can also be applied to the other two parts
of the trajectory leaving the crossing with an angle $\epsilon$ towards the opening of the conductor.
Suppose $t_s$ has a length between $0$ and $T_{{\rm min}}(\epsilon)/2$, then both parts have to be so 
close together that they must leave both through the same lead. As we are interested in the
transmission, we also have to exclude that case\footnote{If we would calculate the reflection instead of
the transmission, the effect of short legs, referred to as coherent backscattering,  has to be taken
into account.}. 
A similar argument holds for the case where the
last part of the orbit has a length between $0$ and $T_{\rm min}(\epsilon)/2$, in this case the orbit has to
come very close to the opening already before the crossing and leave before it could have crossed.
Accounting for all these restrictions, gives the integration limits in Eq.\ (\ref{eq:tra11}).

Now we are prepared to calculate the contribution of the considered trajectory pairs to the transmission.
Therefore we keep in Eq.\ (\ref{eq:tra1}) one sum over trajectories that we will perform using the same 
classical sum rule as in diagonal approximation, the other we can replace by a sum over all the partner 
trajectories  of one trajectory, which can be calculated using $P\left(\epsilon,T\right)$. 
There is, however, one subtlety
concerning the survival probability $\rho(T)$ in the sum rule: we argued already, that if the crossing happens
near the opening, both parts of the orbit act in a correlated way; $\rho(T)$ is changed in the case of the
trajectory pairs considered here for a similar reason: because we know that the two parts of the orbit 
leaving the crossing on each side are very close to each other, the orbit can either leave the cavity 
during the first stretch, \textit{i.e.}\ during the first time it traverses the crossing region, 
or cannot leave at all. This implies that we have to change the survival probability from $\rho(T)$ to
$\rho(T-T_{{\rm min}}(\epsilon))$\footnote{ This effect together with the requirement of a finite length of the
orbit parts leaving towards the opening was originally not taken into account in Ref.\ \cite{Richter02}. In this calculation, the contributions from these two effects cancel each other, they will be only important when considering more complicated diagrams as in the next section.}. Then we arrive at the loop (L) contribution 
{\begin{eqnarray}
\left|t_{nm}(k)\right|_{L}^{2} & = & \frac{\pi\hbar}{2W_{1}W_{2}}\sum_{t}\sum_{P}A_{t}2\Re\exp\left(i\frac{p^{2}\epsilon_{P}^{2}}{2m\lambda\hbar}\right)\nonumber \\
 & = & \frac{4\pi\hbar}{\Sigma(E)}\int_{0}^{\pi}d\epsilon \int_{2T_{{\rm min}}(\epsilon)}^\infty dTe^{-\left( T-T_{{\rm min}}(\epsilon)\right) /\tau_D} P\left(\epsilon,T\right)\cos\left(\frac{p^{2}\epsilon^{2}}{2m\lambda\hbar}\right)\nonumber\\
&=&\frac{8\pi\hbar mv^2\tau_D^3}{\Sigma(E)^2}\int_{0}^{\pi}d\epsilon e^{-T_{{\rm min}}(\epsilon)/\tau_D}\sin\epsilon\cos\left(\frac{p^{2}\epsilon^{2}}{2m\lambda\hbar}\right)
\label{eq:tra24}
\end{eqnarray}
with the sum over the partner trajectories $P$ in the first line. As the important contributions require
very small action differences, \textit{i.e.}\ very similar trajectories, and as the prefactor $A_t$ is not 
as sensitive as the actions to small changes of the trajectories, we can neglect differences between 
$t$ and $t'$ in the prefactor. In the second line we applied the classical sum rule with the modification 
explained before Eq.\ (\ref{eq:tra24}) and used $P\left(\epsilon,T\right)$ to evaluate the sum over $P$. 
After performing the simple time integral in the third line, we can do the $\epsilon$-integration as for 
example in Ref.\ \cite{La�} by taking into account that the important contributions come from very small 
$\epsilon$, yielding
\begin{eqnarray}
\left|t_{nm}(k)\right|_{L}^{2} & = &
\frac{8\pi\hbar mv^2\tau_D^3}{\Sigma(E)^2}\int_{0}^{\pi}d\epsilon(\epsilon/c)^\frac{2}{\lambda\tau_D} \sin\epsilon\cos\left(\frac{p^{2}\epsilon^{2}}{2m\lambda\hbar}\right)\nonumber\\&=&\frac{8\pi\hbar mv^2\tau_D^3}{\Sigma(E)^2} \int dz\frac{m\lambda\hbar}{p^2}\left( \frac{1}{c}\right)^{\left(\frac{2}{\lambda\tau_D} \right)} \left( \frac{2m\lambda\hbar z}{p^2}\right) ^\frac{1}{\lambda\tau_D} \cos z\nonumber\\&=&-\frac{8\pi\hbar mv^2\tau_D^2}{\Sigma(E)^2}\int dz\frac{m\hbar}{p^2}\left( \frac{1}{c}\right)^{\left(\frac{2}{\lambda\tau_D} \right)} \left( \frac{2m\lambda\hbar z}{p^2}\right) ^\frac{1}{\lambda\tau_D} \frac{\sin z}{z}.
\label{epsilo}
\end{eqnarray}
In the first line we already rewrote $e^{-T_{{\rm min}}(\epsilon)/\tau_D}$ as
$(\epsilon/c)^\frac{2}{\lambda\tau_D}$, and in the second line we approximated $\sin\epsilon\approx\epsilon$ and substituted $z=p^2\epsilon^2/\left(2m\lambda\hbar\right)$. Then we perform a partial integration with respect to $z$ neglecting rapidly oscillating terms that are cancelled by the $k$-average, introduced after Eq.\ (\ref{eq:tra1}). Eventually, we perform the $z$-integral by pushing the upper limit to infinity, \textit{i.e.}\ $\hbar\rightarrow 0$ and taking into account our assumption of large dwell times, \textit{i.e.}\ $\lambda \tau_D\rightarrow \infty$. Additionally we assume $\left( 2m\lambda\hbar/p^2\right)^\frac{1}{\lambda\tau_D}\approx 1$; we will return to the last point soon.

Finally, we arrive at the leading nondiagonal contribution to the quantum transmission \cite{Richter02},
\begin{eqnarray}
\left|t_{nm}(k)\right|_{L}^{2} = -\frac{1}{\left(N_{1}+N_{2}\right)^{2}}\, .
\label{eq:tra26.1}\end{eqnarray}

\subsection{Magnetic field dependence of the nondiagonal contribution}

Up to now we assumed time reversal symmetry. If this symmetry is destroyed,  for example  by applying 
a strong magnetic field, the latter contribution will vanish, because the closed loop has to be traversed 
in different directions by the trajectory and its partner. Here we study the transition region between 
zero and finite magnetic field. In particular, we assume a homogeneous magnetic field $B_z$ 
perpendicular to the sample that is assumed weak enough not to change the classical trajectories, but
only the actions in the exponents. Since the closed loop is traversed in different directions by the
two trajectories, we obtain an additional phase difference $\left(4\pi AB_{z}/\phi_{0}\right)$ between 
the two trajectories with the enclosed area $A$ of the loop and the flux quantum $\phi_{0}=\left(hc/e\right)$. 
We further need the distribution of enclosed areas for a trajectory with a closed loop of time $T$ in 
chaotic systems, given by
\begin{equation}
P\left(A,T\right)=\frac{1}{\sqrt{2\pi T\beta}}\exp\left(-\frac{A^{2}}{2T\beta}\right)
\label{eq:tra26.6}
\end{equation}
with a system specific parameter $\beta$. A derivation of this formula can be found for example in Ref.\ \cite{Jen}.
Including the phase difference and the area distribution in a modified $P(\epsilon,T)$ yields
\begin{eqnarray}
P_{B}\left(\epsilon,T\right)\!\!\! & = & \!\!\!\frac{2mv^{2}}{\Sigma(E)}\sin\epsilon\int_{T_{{\rm min}}\left(\epsilon\right)}^{T-T_{{\rm min}}\left(\epsilon\right)}dt_{l}\left(T-t_{l}-T_{{\rm min}}\left(\epsilon\right)\right)\nonumber\\ &&\!\!\!\times\int_{-\infty}^{\infty}dAP\left(A,t_{l}-T_{{\rm min}}\left(\epsilon\right)\right)\cos\frac{4\pi AB_{z}}{\Phi_{0}}\nonumber \\
 & = &\!\!\! \frac{2mv^{2}}{\Sigma(E)}\sin\epsilon\int_{T_{{\rm min}}\left(\epsilon\right)}^{T-T_{{\rm min}}\left(\epsilon\right)}\!\!\!\!dt_{l}\left(T-t_{l}-T_{{\rm min}}\left(\epsilon\right)\right)e^{-\left(t_{l}-T_{{\rm min}}\left(\epsilon\right)\right)/t_{B}}
\label{eq:tra28}
\end{eqnarray}
with $t_{B}=\frac{\phi_{0}^{2}}{8\pi^{2}\beta B_{z}^{2}}$. In the first line we used that paths leaving
the crossing to form a closed loop, enclose a negligible flux, as long as they are correlated; for a more 
detailed analysis see Appendix D of Ref.\ \cite{Jac}. 
Performing the $T$- and $\epsilon$-integrals similar to the case without magnetic field, yields \cite{Richter02}
\begin{eqnarray}
\left|t_{nm}(k,B_z)\right|_{L}^{2} 
  =  -\frac{1}{\left(N_{1}+N_{2}\right)^{2}}\frac{1}{1+\tau_D/t_{B}} \, .
\label{eq:tra30}
\end{eqnarray}
We obtain an inverted Lorentzian with minimum at zero magnetic field, implying that the transmission through 
our sample increases with increasing magnetic field. This weak localization phenomenon, a precursor of
strong localization, is visible as the reduction of the average quantum transmission in Fig.\ \ref{fig:wurm}.

\subsection{Ehrenfest time dependence of the nondiagonal contribution}

This semiclassical approach can also be applied to calculate the Ehrenfest time dependence of the transmission. The Ehrenfest time $\tau_E\equiv(1/\lambda)\ln\left(E/(\lambda\hbar)\right)$, more generally a time proportional to $\ln \hbar$ \cite{ref:Chirikov}, is the time a wave-packet needs to reach a size such that it can no longer be described by a single classical particle. The Ehrenfest time thus separates the evolution of 
wave packets following essentially the classical dynamics (e.g.\ up to few bounces of the wave 
packet in Fig.\ \ref{fig:wavepacket}) from longer time scales dominated by 
wave interference (last panel in Fig.\ \ref{fig:wavepacket}). Based on field theoretical methods, Aleiner
and Larkin showed, that a minimal time is required for quantum effects in the transmission to appear, the Ehrenfest time \cite{ref:Aleiner96}. We will now apply our semiclassical methods to determine the 
Ehrenfest-time dependence of the transmission following the pioneering work \cite{Ada} that has later 
been extended to the reflection \cite{Bro1,Bro2,Jac}, including a distinction between different
Ehrenfest times. To this end, we directly start from Eq.\ (\ref{eq:tra24}); however, we have to be 
more careful when evaluating the $\epsilon$-integral: in our former calculation we assumed 
$e^{-\tau_{\rm E}/\tau_{\rm D}}=\left( \lambda\hbar/E\right)^\frac{1}{\lambda\tau_D} \approx 1$, requiring $\tau_E\ll\tau_D$. Lifting however this strong restriction for $\tau_D$, but still 
keeping it large enough to fulfill our assumption of chaotic dynamics, we obtain
\begin{eqnarray}
\left|t_{nm}(k,\tau_E)\right|_{L}^{2} =   -\frac{1}{\left(N_{1}+N_{2}\right)^{2}}e^{-\tau_{\rm E}/\tau_{\rm D}},
\label{eq:1}
\end{eqnarray}
\textit{i.e.}\ an exponential suppression of the nondiagonal contribution due to the Ehrenfest time. This dependence has been confirmed in numerical simulations for quantum maps.

After this introduction into semiclassical methods for the evaluation of nondiagonal contributions
in configuration space, we will now turn to the generalization to phase space which also allows for
an elegant way to compute higher-order corrections to the leading weak localization contribution
presented above.
          
\section{Quantum transmission: \\
   phase space approach}
\label{sec:phase-space}


The above configuration space treatment, based on self-crossings, is restriced to systems with two
degrees of freedom. More generally, for  higher-dimensional dynamical systems one cannot assume to 
find a one-to-one correspondence between partner orbits and crossings of an orbit \cite{Tur}. In order 
to overcome these difficulties, a phase space approach was developed for calculating the spectral form factor in 
spectral statistics in \cite{Tur,Spe} involving periodic orbits. The next challenge was the 
generalization of this theory to trajectory pairs differing from each other at several places, solved 
again first for the spectral form factor \cite{Heu} and generalized to the transport situation considered 
here in Ref.\ \cite{Mul} which serves as the basis of the following discussion.    

In this section we first explain the phase space approach and use it afterwards in the way developed in Refs.\ \cite{Heu,Mul} for the calculation 
of the quantum transmission involving also higher-order semiclassical diagrams.

\subsection{Phase space approach}

Compared to the last section, we first of all have to replace the role of the reference orbit. Whereas we used there the crossing orbit as reference and calculated then the action difference and the crossing angle distribution in terms of the crossing angle $\epsilon$, we will consider here the orbit without the crossing that is close to itself in the region, where the partner orbit crossed itself in the last section. Inside the region, where the orbit is close to itself; we will refer to it as encounter region and to the parts of the orbit inside as encounter stretches, that are connected by so called links; we place a so called Poincar\'e surface of section $\mathcal{P}$ with its origin at $\tilde{x}=\left(\mathbf{x},\mathbf{p}\right)$. The section consists of all points $\tilde{x}+\delta\tilde{y}=\left(\mathbf{x}+\delta\mathbf{x},\mathbf{p}+\delta\mathbf{p}\right)$ in the same energy shell as the reference point with the $\delta\mathbf{x}$ perpendicular to the momentum $\mathbf{p}$ of the trajectory. For the two-dimensional systems considered here $\mathcal{P}$ is a two-dimensional surface, where every vector $\delta\tilde{y}$ can be expressed in terms of the stable direction $e_{s}\left(\tilde{x}\right)$ and the unstable one $e_{u}\left(\tilde{x}\right)$ \cite{Gas}
\begin{equation}
\delta\tilde{y}=se_{s}\left(\tilde{x}\right)+ue_{u}\left(\tilde{x}\right).
\label{eq:pha1}
\end{equation}
The expressions stable and unstable refer to the following: we consider two orbits, one starting at  $\tilde{x}$ and the other one at $\tilde{x}+\delta\tilde{y}_{0}$, then the difference between the stable coordinates will decrease exponentially for positive times and increase for negative time exponentially in the limit of long time $T$, unstable coordinates  behave just the other way round. The functional form of the exponentials can be determined as $e^{\lambda T}$ and  $e^{-\lambda T}$. Now we can come back to the trajectory with the encounter region, where we have put our Poincar\'e surface of section. The trajectory considered in the last section will pierce through the Poincar\'e section twice: one of this points we will consider as the origin of the section the other piercing will take place at the distance $(s,u)$. The coordinates of the piercing points of the partner trajectory are determined in the following way: the unstable coordinate of the partner trajectory has to be the same as the one of that part of the first trajectory that the second will follow for positive times. The stable coordinate is determined by the same requirement for negative times.

After this introduction into the determination of the $(s,u)$-coordinates, we are now ready to treat trajectory pairs that differ in encounters of arbitrary complexity. Following Ref.\ \cite{Mul} and using the notation introduced there we will allow here that the two trajectories differ in arbitrarily many encounters involving an arbitrary number of stretches. In order to organize this encounter structure we introduce a vector $\vec v=(v_2,v_3,...)$ with the component $v_l$ determining the number of encounters with $l$ stretches involved. The overall number of encounters during an orbit will be denoted by $V=\sum_{l=2}^\infty v_l$, the overall number of encounter stretches by $L=\sum_{l=2}^\infty lv_l$. In an encounter of $l$ stretches, we will get $l-1$ $(s,u)$-coordinates.
 
Now we can proceed by replacing the former expressions for the minimal loop time, the action difference and the crossing angle distribution depending on the former small parameter $\epsilon$, by the corresponding expressions depending on the new small parameters $(s,u)$.

We start with the minimal loop time, to that one also refers as the duration of the encounter. Shifting the Poincar\'e surface of section through our encounter, the stable components will asymptotically decrease, the unstable ones will increase for increasing time. We then claim that both components have to be smaller than a classical constant $c$, its exact value will again, as in the last section, be unimportant for our final results. We finally obtain the encounter duration $t_{{\rm enc}}$ as the sum of the times $t_u$, that the trajectory needs from $\mathcal{P}$ till the point where the first unstable component reaches $c$, and the time $t_s$, that the trajectory needs from $\mathcal{P}$ till the point where the last stable component falls below $c$. Thus we get
\begin{equation}
t_{{\rm enc}}=t_{s}+t_{u}=\frac{1}{\lambda}\ln\frac{c^{2}}{\max_{i}\left\{ \left|s_{i}\right|\right\} \max_{j}\left\{ \left|u_{j}\right|\right\} }.
\label{eq:pha7}
\end{equation}

Now we treat the action difference between the two trajectories. Expressing the actions of the paired trajectories as the line integral of the momentum along the trajectory, we can expand \cite{Tur} one action around the other and express the result in terms of the $(s,u)$-coordinates. This yields for an $l$-encounter\footnote{Strictly speaking \cite{Heu} the $(s,u)$ coordinates used here and in the following calculation are for encounters involving more than two stretches \textit{not} the same like the ones described before, but related to them via a linear and volume preserving transformation.} 
\begin{equation}
\Delta S=\sum_{j=1}^{l-1}s_{j}u_{j}.
\label{eq:pha8}
\end{equation}
The action difference of a trajectory pair is then obtained by adding the differences resulting from all encounters.

Finally we come to the crossing angle distribution that will be replaced here by a weight function for the stable and unstable coordinates for a trajectory of time $T$. We first notice that the uniformity of the trajectory distribution implies in terms of our coordinates in $\mathcal {P}$ that a trajectory pierces through the section with the coordinates $(u,u+du)$ and $(s,s+ds)$ within the time interval $(t,t+dt)$ with the probability $1/\Sigma(E)dsdudt$. In general, we obtain for an $l$-encounter $\left(1/\Sigma(E)\right)^{l-1}ds^{l-1}du^{l-1}dt^{l-1}$. Integrating the product of the latter quantities for all encounters over all possible durations of the $L-V$ intra-encounter links, in a way that their durations are positive yields our weight function for a fixed position of $\mathcal {P}$. To take into account all possible positions of $\mathcal {P}$, we also integrate over all possible positions, where it can be placed and divide by $t_{{\rm enc}}$ to avoid overcounting of equivalent positions. Taking all link times positive, we obtain for the weight function for an orbit of time $T$
\begin{eqnarray}
w_{T}\left({\bf s},{\bf u}\right) \!\!\!&= &\!\!\! \frac{1}{\left(\Sigma\left(E\right)\right)^{L-V}\prod_{\alpha=1}^{V}t_{{\rm enc}}^{\alpha}}\int_{0}^{\infty}dt_{1}\ldots dt_{L}\Theta\left(T-\sum_{\alpha=1}^{V}l_{\alpha}t_{{\rm enc}}^{\alpha}-\sum_{\alpha=1}^Lt_{\alpha}\right)\nonumber \\
 & = &\!\!\! \frac{\left(T-\sum_{\alpha=1}^{V}l_{\alpha}t_{{\rm enc}}^{\alpha}\right)^{L}}{L!\left(\Sigma\left(E\right)\right)^{L-V}\prod_{\alpha=1}^{V}t_{{\rm enc}}^{\alpha}}.
\label{eq:pha9}
\end{eqnarray}

One additional problem arises, when treating trajectory pairs differing not only in one $2$-encounter like in the last section: one can construct for one $\vec v$ different trajectory pairs, varying for example in the relative orientation, in  which the encounter stretches are traversed. We will count this number by a function $N\left(\vec v\right)$ and describe briefly later, how it can be calculated. 

\subsection{Calculation of the full transmission}

After the introduction into the phase space approach we are now ready to calculate the transmission. Taking the weight function, the action difference and the number of structures, we can transform the non-diagonal (ND) part of Eq.\ (\ref{eq:tra1}) into
\begin{eqnarray}
\left|t_{nm}(k)\right|_{ND}^{2} & =& \frac{\pi\hbar}{2W_{1}W_{2}}\sum_{t}\sum_{\vec{v}}A_tN\left(\vec{v}\right)\nonumber \\
 &  & \times\left\langle \int_{-c}^{c}...\int_{-c}^{c}d^{L-V}ud^{L-V}s\:\exp\left(\frac{i}{\hbar}\Delta S\right)w_{T}\left(\textbf{s},\textbf{u}\right)\right\rangle _{\Delta k}
\label{eq:pha9.7}
\end{eqnarray}
with the average over a small $k$-window denoted by $\left\langle\ldots \right\rangle_{\Delta k}$. Inserting the formulas for the action difference, the weight function and using the classical sum rule with the modification of the survival probability, discussed in the last section, we can then transform the integral with respect to the length of the trajectory into one over the last link and obtain
\begin{eqnarray}
\left|t_{nm}(k)\right|_{ND}^{2} & = & \frac{2\pi\hbar}{\Sigma(E)}\sum_{\vec{v}}N\left(\vec{v}\right)\left(\prod_{i=1}^{L+1}\int_{0}^{\infty}dt_{i}\exp\left(-\frac{t_{i}}{\tau_D}\right)\right)\nonumber \\
 &  &\hspace*{-1.07cm} \times\left\langle \int_{-c}^{c}...\int_{-c}^{c}\frac{d^{L-V}ud^{L-V}s}{\left(\Sigma\left(E\right)\right)^{L-V}}\prod_{\alpha=1}^{V}\frac{\exp\left(-\frac{t_{{\rm enc}}^{\alpha}}{\tau_D}+\frac{i}{\hbar}\sum_{j=1}^{l_{\alpha}-1}s_{\alpha j}u_{\alpha j}\right)}{t_{{\rm enc}}^{\alpha}}\right\rangle _{\Delta k}\nonumber \\
 & =& \frac{1}{N_{1}+N_{2}}\sum_{n=1}^{\infty}\left(\frac{1}{N_{1}+N_{2}}\right)^{n}\quad\sum_{\vec{v}}^{L-V=n}\left(-1\right)^{V}N\left(\vec{v}\right)
\label{eq:pha10}
\end{eqnarray}
with the $L+1$ link times $t_i$. The $(s,u)$-integrals are calculated using the rule \cite{Heu}, that after 
expanding the exponential $e^{t_{{\rm enc}}/\tau_D}$ into a Taylor series, only the $t_{{\rm enc}}$-independent term contributes and yields in leading order in $\hbar$
\begin{equation}
\left\langle \frac{1}{\Sigma\left(E\right)}\int_{-c}^{c}\int_{-c}^{c}dsdu\exp\left(\frac{isu}{\hbar}\right)\right\rangle_{\Delta k} \sim\frac{1}{T_{H}}
\label{eq:pha11}
\end{equation}
with the so called Heisenberg time $T_H=\Sigma(E)/(2\pi\hbar)$. For the sum with respect to $\vec{v}$, one can derive recursion relations, yielding \cite{Heu} 
\begin{equation}
\sum_{\vec{v}}^{L-V=n}\left(-1\right)^{V}N\left(\vec{v}\right)=\left(1-\frac{2}{\beta}\right)^{n}
\label{eq:pha12}
\end{equation} 
with $\beta=1$ and $\beta=2$ for the case with and without time reversal symmetry, respectively. Relations of this kind are derived by describing our trajectories by permutations expressing the connections inside and between encounters and considering the effect of shrinking one link in an arbitrarily complicated structure to zero.

We then obtain for $T(E_F)$, given in Eq.\ (\ref{Landauer}), in the case with time reversal symmetry
\cite{Mul}
\begin{equation}
T(E_F)^{\beta=1}\approx\frac{N_{1}N_{2}}{N_{1}+N_{2}}+\frac{N_{1}N_{2}}{N_{1}+N_{2}}\sum_{n=1}^{\infty}\left(\frac{-1}{N_{1}+N_{2}}\right)^{n}=\frac{N_{1}N_{2}}{N_{1}+N_{2}+1}
\label{eq:pha14}
\end{equation}
and in the case without time reversal symmetry
\begin{equation}
T(E_F)^{\beta=2}\approx\frac{N_{1}N_{2}}{N_{1}+N_{2}},
\label{eq:pha15}
\end{equation}
which  agrees with the diagonal contribution, already obtained in the previous section. 
Both results are in agreement with  RMT predictions \cite{Bee97}.

\section{Semiclassical research paths: \\
  present and future}

In the above sections we outlined the recently developed semiclassical techniques
for treating off-diagonal contributions for one paradigmatic example of coherent
transport, weak localization. Nonlinear (hyperbolic) dynamics as a prerequisite for 
the formation of correlated orbit pairs, together with classical ergodicity are at the core 
of the universality in the interference contribution to the ballistic conductance. 
Recently, this semiclassical approach to  quantum transport has been extended
along various paths, thereby gaining a more and more closed semiclassical framework
of mesoscopic quantum effects:

In the presence of spin-orbit interaction, spin relaxation in confined ballistic systems 
and weak antilocalization, the \textit{enhancement} of the conductance for systems obeying 
time-reversal symmetry (\textit{i.e.}\ at zero magnetic field), has been semiclassically 
predicted \cite{Zaitsev-et-al-prl}
and generalized to other types of spin-orbit interaction \cite{Zaitsev-et-al-prb}
and higher-order contributions \cite{waltner-bolte-prb}.

Ballistic conductance fluctuations, the analogue of the universal conductance
fluctuations in the diffusive case, require the computation of semiclassical 
diagrams involving four trajectories. Again, RMT predictions could be
confirmed \cite{Bro2}. Accordingly, RMT predictions for shot noise in Landauer transport agree with recent
semiclassical results \cite{Whit}. This could also be shown for higher moments of the conductance in \cite{Nov}.
Even strong localization effects (in certain systems, {\em i.e.}\ chains of chaotic 
cavities) could be derived by making use of semiclassical loop-contributions \cite{BrouwerXY}.

Finally, beyond RMT, Ehrenfest time effects on transport properties have recently been
in the focus of the field. As a result, ballistic conductance fluctuations turn out to be, 
to leading order,
$\tau_{\rm E}$-independent \cite{Bro1}, contrary to weak localization which 
is suppressed with $e^{-\tau_{\rm E} / \tau_{\rm D}}$ \cite{Ada}, as we have shown.
Recently these results were extended to the ac-conductance in \cite{Pet}.

Beside universal spectral statistics of closed systems and transport through open
systems, the semiclassical techniques have been recently generalized to decay and
photo fragmentation of complex systems \cite{Waltner-et-al-prl}. 

To summarize, by now there exists a newly developed semiclassical machinery to compute 
systematically quantum coherence effects for quantities being composed of  products of
Green functions. This opens up various 
possible future research directions: With respect to system classes, mesoscopic
transport theory
has been predominantly focused on electron transport, leaving the conductance
of ballistic hole systems aside, which plays experimentally an important role and
for which a semiclassical theory remains to be developed. The situation is similar 
for a novel class of ballistic conductors: graphene based nanostructures,
see Fig.\ \ref{fig:wurm}. More generally, a proper treatment of interaction effects 
in mesoscopic physics would probably represent the major challenge to future semiclassical 
theory.

To conclude, at mesoscopic scales where linear quantum evolution meets non-linear classical
dynamics, further interesting phenomena can be expected to emerge, in view of increasingly
controllable future experiments.

\end{document}